\documentclass[aps,prb,reprint,twocolumn,showpacs,preprintnumbers,amsmath,amssymb,floatfix]{revtex4}

\usepackage{graphicx}
\usepackage{dcolumn}
\usepackage{bm}
\usepackage{dsfont}

\begin{document}

\title{Nuclear Spins of Ionized Phosphorus Donors in Silicon}

\author{Lukas Dreher}
\email{dreher@wsi.tum.de} 
\author{Felix Hoehne}
\email{hoehne@wsi.tum.de} 
\author{Martin Stutzmann}
\author{Martin S. Brandt}
\affiliation{Walter Schottky Institut, Technische Universit\"at M\"unchen, Am Coulombwall 4, 85748 Garching, Germany}


\date{\today}
\begin{abstract}
We demonstrate the coherent control and electrical readout of ionized phosphorus donor nuclear spins in $^{\mathrm{nat}}$Si. By combining time-programed optical excitation with coherent electron spin manipulation, we selectively ionize the donors depending on their nuclear spin state, exploiting a spin-dependent recombination process at the Si/SiO$_2$ interface, and find a nuclear spin coherence time of 18~ms for the ionized donors. The presented technique allows for spectroscopy of ionized-donor nuclear spins and enhances the sensitivity of electron nuclear double resonance to a level of 3000 nuclear spins.
\end{abstract}

\maketitle
Nuclear spins in semiconductors are well isolated quantum systems and therefore excellent candidates for a quantum memory \cite{N455_1085,NPhys_Fuchs}. The hyperfine coupling between a nuclear spin and an electron spin residing in its vicinity makes the nuclear spin state accessible to optical and electrical readout schemes \cite{PRL93_130501,S330_1652,PRL106_187601,JoAP109_102411,NPhys_Fuchs}. It allows the transfer of spin coherence between the electron and nucleus \cite{N455_1085,NPhys_Fuchs} and is a prerequisite for the widely used electron nuclear double resonance (ENDOR) technique \cite{PR114_1219}. On the other hand, this hyperfine interaction also couples the nuclear spin to its environment, which can result in a loss of coherence and leads to a broadening of resonance lines \cite{Kohei_Spintech}, possibly hiding spectroscopic detail. It would therefore be desirable to switch on and off the hyperfine interaction, e.g., by controlling the charge state of a donor. In the context of silicon-based quantum computing \cite{N393_133}, the electron and nuclear spins of phosphorus donors in the neutral charge state have been studied extensively; spin coherence times of up to seconds and minutes have been reported for the electron and nuclear spin in Si highly enriched with $^{28}$Si, respectively \cite{Tyryshkin_10sec,Thewalt_Spintech}. Significant progress in the fabrication of few-donor devices \cite{APL86_202101,NN5_502} has enabled the single-shot readout of a single $^{31} $P electron spin \cite{N467_687}, and the electrical readout of $^{31}$P nuclear spins has been achieved by using pulsed electrically detected ENDOR (EDENDOR) \cite{S330_1652,PRL106_187601,mccamey_electrically_2011}. 
While these EDENDOR studies have addressed the spectroscopy and dynamics of the $^{31}$P nuclear spin in the neutral donor state ($^{31}$P$^0_\mathrm{n}$), the dynamics of the ionized-donor nuclear spin ($^{31}$P$^+_\mathrm{n}$) has so far been unexplored.

Combining EDENDOR with time-programed optical excitation, we here demonstrate that the $^{31}$P donors can be selectively depopulated (i.e. ionized) depending on the orientation of their nuclear spin. This makes it possible to manipulate and electrically read out the nuclear spins of the ionized donors. We show that the coherence time of the $^{31}$P$^+_\mathrm{n}$ in $^{\mathrm{nat}}$Si is increased by 2 orders of magnitude with respect to the corresponding $^{31}$P$^0_\mathrm{n}$ in the structures studied, rendering the $^{31}$P$^+_\mathrm{n}$ a possible resource for a quantum spin memory, particularly in devices where the donor resides close to an interface. Furthermore, the selective depopulation scheme employed here enables EDENDOR spectroscopy with a sensitivity of $<3000$ nuclear spins, orders of magnitude more sensitive than in previous experiments \cite{PRL106_187601}.

We used a [001]-oriented Si:P silicon-on-insulator sample, where the top 20~nm were phosphorus-doped ($[\mathrm{P}]=3\times 10^{16}$~cm$^{-3}$). The sample was placed in an external magnetic field of $B_0=0.3503$~T ($B_0||[110]$) at 5~K in a dielectric microwave resonator for pulsed ENDOR. It was illuminated with the light of a pulsed LED (Thorlabs LDC 210 controller) with a rise time of $\approx 1\mu$s and a wavelength of 625~nm at an intensity of 20 mW/cm$^2$. The photocurrent through the sample ($\approx 22~\mu$A) was measured under symmetric bias (300~mV) by using a balanced transimpedance amplifier with low- and high-pass filtering at cut-off frequencies of 1~MHz and 2~kHz, respectively. The microwave (mw) frequency was set to be in resonance with the high-field resonance of the hyperfine-split $^{31}$P$_\mathrm{e}$ transition \cite{PRB83_235201}. For noise reduction, a lock-in detection scheme was employed \cite{PRL106_187601,arXiv_1111_5149}. Further details of the experimental techniques can be found in Refs.~\cite{PRL106_187601,PRB83_235201,arXiv_1111_5149} and in the appendix \ref{Appendix_A}. 

The electrical nuclear spin readout is based on a spin-dependent recombination process via weakly coupled spin pairs \cite{NP2_835} formed by $^{31}$P donor electron spins ($^{31}$P$_\mathrm{e}$) and paramagnetic dangling bond states P$_\mathrm{b0}$ \cite{JAP83_2449} at the Si/SiO$_2$ interface \cite{JdP39_51,PRL104_46402}. Crucial for the selective depopulation scheme of the $^{31}$P donors is the fact that, due to the Pauli principle, the lifetime of the parallel $^{31}$P$_\mathrm{e}$-P$_\mathrm{b0}$ spin-pair states is substantially longer than that of the antiparallel pairs. In a first experiment, we therefore experimentally determine these lifetimes.
\begin{figure}[]
\includegraphics[]{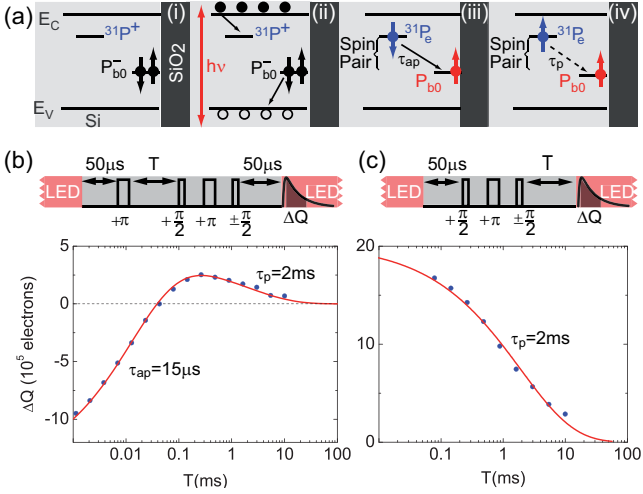}
\caption{(color online)(a) Recombination involving the $^{31}$P$_\mathrm{e}$-P$_\mathrm{b0}$ spin pair: diamagnetic $^{31}$P$^+$-P$_\mathrm{b0}^-$ state, the ground state in the dark (i). Formation of spin pairs by optical excitation of charge carriers, capture, and recombination (ii). The spin-dependent transitions are characterized by the time constants $\tau_\mathrm{ap}$ for antiparallel (iii) and $\tau_\mathrm{p}$ for parallel (iv) configuration of the $^{31}$P$_\mathrm{e}$ and P$_\mathrm{b0}$ spins. (b) Inversion recovery pulse sequence with pulsed illumination  to experimentally determine $\tau_\mathrm{ap}$; boxcar integrating the current transient yields $\Delta Q$. The echo signal $\Delta Q$ (solid circles) decays for short $T$ with a time constant $\tau_\mathrm{ap}=15~\mu$s. (c) Pulse sequence for the determination of $\tau_\mathrm{p}$. $\Delta Q$ decays as a function of the waiting time $T$ between the spin echo and the LED pulse (solid circles) with a characteristic time constant $\tau_\mathrm{p}=2$~ms.}
\label{fig:ElectronicTimeScales}
\end{figure}
We discuss the dynamics of the spin pair in terms of the model sketched in Fig.~\ref{fig:ElectronicTimeScales} (a) \cite{JdP39_51,PRB68_245105}. We assume that, without illumination, the $^{31}$P donors at the Si/SiO$_2$ interface are compensated by interface defects and therefore are in the ionized $^{31}$P$^{+}$ state as sketched in panel (i).  Upon illumination (ii), the donors become occupied forming $^{31}$P$_\mathrm{e}$-P$_\mathrm{b0}$ spin pairs (iii) and (iv).
The spin pair will return to the $^{31}$P$^+$-P$_\mathrm{b0}^-$ state (i) on a time scale of $\tau_\mathrm{ap}$ for antiparallel spin configuration (iii) or remain stable on much longer time scales $\tau_\mathrm{p}$ for parallel spin orientation (iv). 
Consequently, a dynamic equilibrium is established, in which in good approximation all of the spin pairs are in the parallel configuration, which we refer to as ``the steady state.''
To determine $\tau_\mathrm{ap}$, we employ the pulse sequence shown in Fig.~\ref{fig:ElectronicTimeScales} (b), resembling an inversion recovery experiment \cite{Schweiger_Book,PRB81_75214}.
After switching off the LED, a mw $\pi$-pulse inverts the $^{31}$P$_\mathrm{e}$, bringing the spin pairs into an antiparallel configuration. 
A detection echo is applied after a waiting time $T$, followed by a second LED pulse, during which the photocurrent is recorded and boxcar integrated. This  results in a charge $\Delta Q$, which can be shown to be proportional to the difference of the number of parallel and antiparallel spin pairs as described in the appendix \ref{Appendix_B}. An electrical readout scheme including a pulsed laser flash has conceptually been discussed e.g., in Ref.~\cite{pss233_427}.
As shown in Fig.~\ref{fig:ElectronicTimeScales} (b), the signal decays as a function of $T$ (solid circles). The decay is described by a sum of two stretched exponentials ($\exp[-(T/\tau)^n]$) with $\tau_\mathrm{ap}=15~\mu$s, $\tau_\mathrm{p}=2$~ms, and $n=0.5$ for both exponentials, reflecting the distribution of $^{31}$P-P$_\mathrm{b0}$ distances. We identify the fast time constant with the transition time $\tau_\mathrm{ap}$ to the $^{31}$P$^+$-P$_\mathrm{b0}^-$ state. The fact that for large $T$ the signal decays to zero indicates that almost all of the $^{31}$P contributing to the spin-dependent signal are in the unoccupied $^{31}$P$^{+}$ state in the absence of illumination so that we have control over the $^{31}$P charge state.  Because of nonidealities of the mw $\pi$ pulse, some of the $^{31}$P$_\mathrm{e}$ are not inverted, and thus the corresponding spin pairs remain in parallel configuration. Additionally, the finite lifetime of conduction band electrons leads to the formation of new spin pairs even after the LED has been switched off. This results in the positive $\Delta Q$ for $T>0.2$~ms which decays with a time constant of 2~ms, identified with $\tau_\mathrm{p}$. This time constant can also be accessed directly in a separate experiment, where the time interval between an electron spin echo and the light pulse is varied as shown in Fig.\ref{fig:ElectronicTimeScales}~(c). For $T\gg\tau_\mathrm{ap}$, the signal intensity is proportional to the spin pairs in the parallel state after the detection echo, thus allowing a second measurement of $\tau_\mathrm{p}$ as discussed in the appendix \ref{Appendix_B}. The observed signal decay can be described by a stretched exponential with $\tau_\mathrm{p}=2$~ms and $n=0.5$, which were used as fixed parameters in the fit in Fig.~\ref{fig:ElectronicTimeScales} (b).

\begin{figure}[]
\includegraphics[]{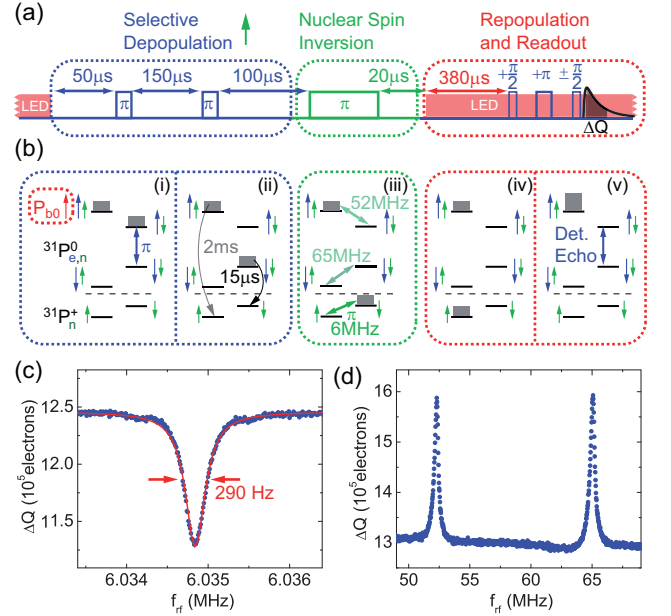}
\caption{(color online) (a) Pulse sequence for the formation, manipulation, and electrical detection of the $^{31}$P$^+_\mathrm{n}$. (b) The $^{31}$P electron and nuclear spins are represented by blue (large) and green (small) arrows, respectively.  We draw the four hyperfine-split levels of the occupied donor in the upper part of each panel and, separated by a dashed line, the levels of the ionized-donor nuclear spin in the lower part. The populations of the levels throughout the pulse sequence are indicated by gray boxes. For simplicity, we depict a subensemble of spin pairs with the P$_\mathrm{b0}$ spin in the spin-up state, indicated by the red arrow in panel (i). In (c) and (d), the detection echo amplitude $\Delta Q$ is shown as a function of the rf pulse frequency $f_\mathrm{rf}$, revealing a quenching and enhancement of $\Delta Q$, when the frequency is resonant with the transitions labeled by 6, 52, and 65~MHz in (iii), respectively.}
\label{fig:SelectiveDepopulation}
\end{figure}
Having established the dynamics of the spin pair, we devise the scheme for the manipulation and readout of the ionized $^{31}$P$^+_\mathrm{n}$ state illustrated in Fig.~\ref{fig:SelectiveDepopulation} (a) and (b). We sketch the four energy levels of the hyperfine-split occupied $^{31}$P donor ($S=1/2$, $I=1/2$) with the corresponding electron and nuclear spin states shown in the upper part of panels (i)-(v) and the $^{31}$P$^+_\mathrm{n}$ levels shown in the lower part separated by a dashed line. For simplicity, we show only the $^{31}$P$_\mathrm{e}$-P$_\mathrm{b0}$ subensemble with the P$_\mathrm{b0}$ spin in the  ``spin-up'' state, indicated by the red arrow in (i). 
At the beginning of the pulse sequence, the spin pairs are in the steady state (i). Note that for the subensemble with the P$_\mathrm{b0}$ spin in the ``spin-down'' state, the populations are reversed when compared to panel (i), and therefore there is no net polarization of the spin system, neglecting the thermal equilibrium polarization. A mw $\pi$ pulse resonant with one of the $^{31}$P$_\mathrm{e}$ hyperfine transitions converts the electron spin pairs associated with one $^{31}$P nuclear spin state into antiparallel configuration. Thus, the donors with this nuclear spin state become ionized on the time scale of $\tau_\mathrm{ap}$ (ii). To compensate for imperfections of the first depopulation pulse, we apply an additional depopulation pulse separated by 150~$\mu$s to also ionize the remaining donors with this nuclear spin state.
This selective depopulation scheme results in a large population difference of the $^{31}$P$^+_\mathrm{n}$ levels as shown in (iii), allowing for manipulation and readout of the $^{31}$P$^+_\mathrm{n}$. Application of a radio frequency (rf) $\pi$ pulse with a frequency of $f_\mathrm{rf}\approx$6~MHz (iii) inverts the populations of the $^{31}$P$^+_\mathrm{n}$, creating a nuclear spin polarization exceeding the thermal equilibrium polarization (iv). After switching on the LED, the ionized donors become repopulated and the steady state of the electronic system is established. We assume that the repopulation process does not change the states of the nuclear spins, resulting in a nuclear spin polarization of the occupied donors (v). For a nonresonant rf pulse, the level populations at the end of the pulse sequence are identical to the ones shown in (i). The difference between the spin populations  on the corresponding hyperfine transition in the resonant and nonresonant cases can be quantified by measuring the amplitude $\Delta Q$ of a detection echo with phase cycling \cite{PRL106_187601} (cf.~appendix \ref{Appendix_C}), indicated by the blue arrow in (v). In addition, the population differences of the $^{31}$P$^0_\mathrm{n}$ levels prevail on the time scale of $\tau_\mathrm{p}=2$~ms, orders of magnitudes longer than in previous EDENDOR experiments \cite{PRL106_187601}, where the manipulation of the nuclear spins was limited by $\tau_\mathrm{ap}\ll\tau_\mathrm{p}$, thus enabling
also improved experiments on the  $^{31}$P$^0_\mathrm{n}$.

In Figs.~\ref{fig:SelectiveDepopulation} (c) and (d), $\Delta Q$ is shown as a function of the rf pulse frequency, revealing a quenching of the echo signal at a nuclear spin resonance frequency of 6.034\,84(1)~MHz and an enhancement at the frequencies of 52.279(3) and 65.042(3)~MHz. A quenching of the echo signal is expected for a resonant transition of the $^{31}$P$^+_\mathrm{n}$ when considering the population differences for the corresponding hyperfine transition shown in (i) and (v). For the $^{31}$P$^0_\mathrm{n}$ the enhancement of the echo signal can also be understood in terms of Fig.~\ref{fig:SelectiveDepopulation} when the populations are inverted, e.g., on the 52~MHz transition instead of the 6~MHz transition; cf. panel (iii). From the resonance frequency of the $^{31}$P$^+_\mathrm{n}$, we extract a nuclear $g$ factor of $g_\mathrm{n}=-2.2601(3)$; this corresponds to a chemical shift of -1400(150)~ppm relative to the free nucleus \cite{ADaNDT90_75}, assuming an uncertainty of $\pm$0.05~mT in $B_0$. The chemical shift of the $^{31}$P$^+_\mathrm{n}$ relative to the $^{31}$P$^0_\mathrm{n}$ is 710(10)~ppm, which can be determined more precisely, since it is affected to a lesser extent by a systematic error in $B_0$. 

In conventional ENDOR experiments of partially compensated phosphorus-doped silicon, a resonance approximately at the free $^{31}$P$_\mathrm{n}$ Larmor frequency has been observed and attributed to $^{31}$P$^+_\mathrm{n}$ weakly hyperfine-coupled to neighboring isolated $^{31}$P$^0$ \cite{PR114_1219} or $^{31}$P clusters at higher $^{31}$P concentrations \cite{PR134_1001a}. While we cannot completely rule out a contribution to the observed signal through such a mechanism, it seems unlikely given the nonselectivity of the here employed Davies-type of ENDOR with respect to small hyperfine interactions \cite{Schweiger_Book} and the low $^{31}$P concentration of the sample studied. Also, the doubly occupied donor state $^{31}$P$^-$ in its singlet electron spin state is expected to exhibit a nuclear Larmor frequency near that of the free nucleus. While in high magnetic field EDENDOR experiments \cite{S330_1652} the $^{31}$P$^-$ state is thought to be involved in the $^{31}$P$^0_\mathrm{n}$ readout, at the magnetic field and temperature used in this work the $^{31}$P-P$_\mathrm{b0}$ recombination is the dominant spin-dependent process \cite{PRL104_46402}. We therefore attribute the observed resonance at 6~MHz to the nuclear spins of the donors selectively ionized with the mechanism described in Fig.~\ref{fig:SelectiveDepopulation} (a).

From the data in Fig.~\ref{fig:SelectiveDepopulation} (d), we infer a signal-to-noise ratio of $S/N\approx 100$ and a sensitivity of $<3000$ nuclear spins for a measurement time of $\approx$40~min. This nuclear spin sensitivity was determined from the noise in $\Delta Q$ under the assumption that one nuclear spin corresponds to one electronic charge. In comparison with the EDENDOR spectroscopy data shown in Ref.~\cite{PRL106_187601}, the $S/N$ is improved by more than 2 orders of magnitude for comparable measurement times and the pronounced nonresonant background is almost entirely removed. 

\begin{figure}[]
\includegraphics[]{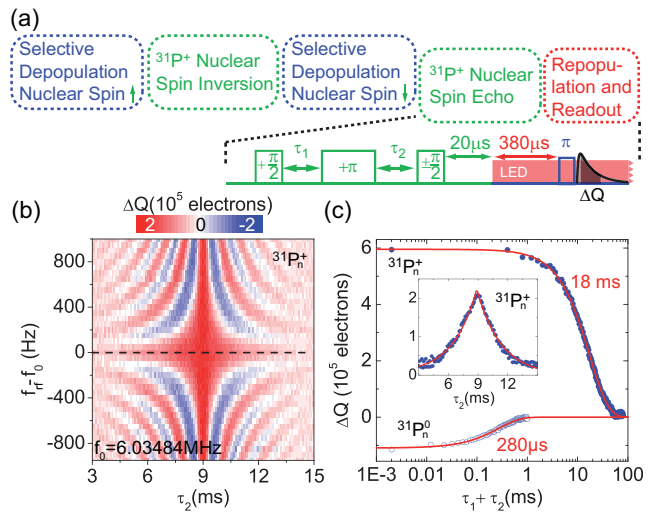}
\caption{(color online) (a) The pulse sequence for the electrically detected nuclear spin echo of the $^{31}$P$^+_\mathrm{n}$ is outlined by schematic building blocks, representing pulse sequences depicted in detail in Fig.~\ref{fig:SelectiveDepopulation}. (b) shows the nuclear spin echo amplitude as a function of the rf pulse frequency and $\tau_2$ for $\tau_1=9$~ms in a contour plot. In (c) nuclear spin echo decays of the $^{31}$P$^+_\mathrm{n}$ (dots) and of the $^{31}$P$^0_\mathrm{n}$ (open circles) are shown together with stretched exponential fits (lines), revealing the indicated coherence times. The inset shows the $^{31}$P$^+_\mathrm{n}$ echo fitted by two back-to-back exponential decays (line).}
\label{fig:DynamicsIonizedDonor}
\end{figure}

To investigate the dynamics of the $^{31}$P$^+_\mathrm{n}$, we prepare the nuclear spin system as described in Fig.~\ref{fig:DynamicsIonizedDonor} (a). We selectively depopulate the levels associated with one nuclear spin state and invert the populations of the $^{31}$P$^+_\mathrm{n}$ levels, resulting in the level populations shown in Fig.~\ref{fig:SelectiveDepopulation} (iv). Subsequently, also the levels associated with the other nuclear spin state are depopulated, further enlarging the population difference of the $^{31}$P$^+_\mathrm{n}$ levels. This population difference is expected to persist on the time scale of the nuclear spin lifetime, allowing us to measure the dynamics of the $^{31}$P$^+_\mathrm{n}$ on a time scale much longer than $\tau_\mathrm{p}=2~$ms.
We employ the nuclear spin echo pulse sequence shown in Fig.~\ref{fig:DynamicsIonizedDonor} (a). Since the readout preserves nuclear spin populations but not coherence, we project the nuclear spin system into one of its eigenstates with the final rf $\pi/2$ pulse, and the population difference of the eigenstates is read out after repopulating the $^{31}$P$^0$ levels. In the case of the nuclear spin echo of the $^{31}$P$^+_\mathrm{n}$ experiments, the readout consists of a single mw $\pi$ pulse. Lock-in detection is realized by cycling the phase of the final rf $\pi/2$ pulse by 180$^\circ$ from shot to shot; cf.~appendix \ref{Appendix_C}.
Figure \ref{fig:DynamicsIonizedDonor} (b) shows the electrically detected nuclear spin echo amplitude in a contour color plot for $\tau_1=9$~ms as a function of the rf pulse frequency and $\tau_2$. At the resonance frequency $f_0=6.034\,84$~MHz, the nuclear spin echo, shown in the inset in Fig.~\ref{fig:DynamicsIonizedDonor} (c), is fitted by two back-to-back exponential decays \cite{Schweiger_Book} with a time constant of $\tau$=2.4~ms, corresponding to a linewidth of $1/(\pi \tau)$=132~Hz in the frequency domain. This is roughly a factor of 2 smaller than the value extracted from the linewidth of the dip in Fig.~\ref{fig:SelectiveDepopulation} (c), which is spectrally broadened by the rf pulse used there. For off-resonant frequencies, the signal oscillates as a function of $\tau_2$ resulting in the characteristic pattern of the contour plot, which can be quantitatively modeled by a matrix formalism for nuclear induction \cite{PR98_1099,PRB83_235201}.

To determine the coherence time of the $^{31}$P$^+_\mathrm{n}$, we measure the nuclear spin echo amplitude as a function of $\tau_1=\tau_2$. The solid circles in Fig.~\ref{fig:DynamicsIonizedDonor} (c) show the resulting signal $\Delta Q$ as a function of $\tau_1+\tau_2$, revealing a stretched exponential decay with a time constant of 18~ms and an exponent of 1.2. For comparison, the open circles show the nuclear spin echo decay of the $^{31}$P$^0_\mathrm{n}$, measured on the 52~MHz transition after depopulating only the levels associated with one nuclear spin orientation. An exponential fit reveals a coherence time of 280~$\mu$s, almost 2 orders of magnitude smaller than the 18~ms obtained for the $^{31}$P$^+_\mathrm{n}$. This shows the potential benefit of using the $^{31}$P$^+_\mathrm{n}$ as quantum spin memory \cite{N455_1085}.

The observed $^{31}$P$^0_\mathrm{n}$ coherence time is shorter than the one measured in bulk natural silicon \cite{mccamey_electrically_2011,Tyryshkin_Lyon_private}, suggesting that in our sample additional decoherence of the $^{31}$P$^0_\mathrm{n}$ is caused by the presence of the nearby interface, e.g., via the hyperfine interaction by the coupling of the donor electron to fluctuating charges in the oxide.

In the case of the $^{31}$P$^+_\mathrm{n}$, this electron-mediated coupling is absent, resulting in the much longer coherence time of 18~ms observed in the presented experiment. Calculations show that spectral diffusion due to dipolar coupling of the $^{31}$P$^+_\mathrm{n}$ to $^{29}$Si nuclear spins results in a stretched exponential echo decay with a coherence time of $\approx$30~ms and an exponent of $\approx$1.4 \cite{WayneWitzel_private}, in good agreement with our experimental results. A further possible mechanism of decoherence of the $^{31}$P$^+_\mathrm{n}$ are fluctuating magnetic fields generated by spin flips of interface defects \cite{PRB76_245306}.

In summary, we have used pulsed illumination in combination with coherent spin manipulation to selectively depopulate the $^{31}$P donors depending on their nuclear spin state. In the field of electrically detected magnetic resonance (EDMR), this combination allows us to experimentally access parameters involved in the spin-dependent transport process \cite{PRB68_245105} such as the recombination time of parallel spin pairs. Based on the selective depopulation technique, we have achieved spectroscopy of ionized-donor nuclear spins and investigated their coherence time by means of EDENDOR. Since the linewidth of the nuclear spin transition can be greatly reduced by ionizing the donor, this technique allows for a more precise spectroscopy of the nuclear spin, e.g. by studying the influence of local strains \cite{PRL106_37601} or electric fields on the $^{31}$P$^+_\mathrm{n}$ resonance. This method can be applied to other donors in silicon with $I>1/2$, e.g.~$^{209}$Bi, where the investigation of the nuclear quadrupole splitting as a function of strain is an appealing challenge~\cite{APL96_32102}.  Furthermore, the achieved $S/N$ is orders of magnitude larger compared to previous EDENDOR experiments \cite{PRL106_187601}. This makes the presented selective depopulation scheme particularly useful for studying defects in semiconductor nanostructures. The longer coherence time of the $^{31}$P$^+_\mathrm{n}$ compared to the $^{31}$P$^0_\mathrm{n}$ renders the ionized donor an attractive candidate for a quantum spin memory \cite{N455_1085}. To realize such a quantum memory, the nuclear spin coherence must not be destroyed by the process of depopulating and repopulating the donor. Therefore, the ionization and deionization should take place deterministically, which could be realized, e.g., by electric gates \cite{NL7_2000} or optical excitation \cite{PRL102_257401}, rather than by the statistical recombination process employed here. Such coherence-preserving ionization schemes are also of interest in the context of cluster state quantum computing \cite{Morton_Cluster}.

We thank Wayne M.~Witzel, Alexei~M.~Tyryshkin, and Stephen A.~Lyon for fruitful discussions. The work was supported by DFG (Grant No. SFB 631, C3) and by BMBF (EPR-Solar).

\appendix
%

\section{Experimental Details}\label{Appendix_A}

The microwave (mw) pulses and the radio frequency (rf) pulses were amplified by a traveling wave tube amplifier and a 300~W solid state amplifier, respectively. The mw power level was adjusted such that the $\pi$-pulse length was 40~ns. In the experiments shown in Fig.~1 of the main text, the shot repetition time (SRT) was set to 21.4~ms; the waiting time between the $\pi$ pulse and the $\pi/2$ pulses of the detection echo was 100~ns, as in all electron-spin echoes employed in this work. The data shown in Fig.~2 (c) of the main text were taken at SRT=11.4~ms and an rf pulse length of $T_\mathrm{rf}=10$~ms with the rf amplifier attenuated such that the rf $B_2$-field was 150~nT; the data in Fig.~2 (d) of the main text were recorded with SRT=1.2~ms, $T_\mathrm{rf}=10\mu$s and $B_2=150~\mu$T. The rf $\pi$-pulse length used for measuring the $^{31}$P$^+_\mathrm{n}$ echoes was $T_\pi=19~\mu$s, as determined by measuring coherent nuclear spin oscillations, and the SRT was 97~ms. For the $^{31}$P$^0_\mathrm{n}$ echo measurements, the SRT was 9.7~ms. In this case, a capacitive network was used to match the impedance of the rf amplifier to the rf coil such that the $\pi$-pulse time could be reduced to $T_\pi=4~\mu$s, allowing to excite a larger fraction of the resonance line \cite{PRL106_187601}.

\section{Spin-to-charge conversion under pulsed illumination}\label{Appendix_B}

\begin{figure}[h]
\includegraphics[]{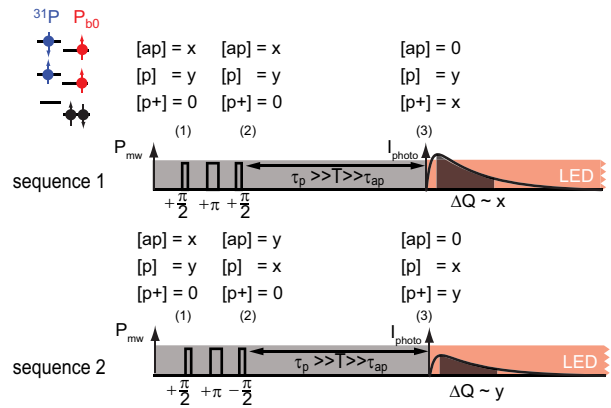}
\caption{Readout pulse sequence resulting in a charge $\Delta Q$ proportional to the difference between the number of antiparallel and parallel spin pairs at the beginning of the readout pulse sequence. The number of spin pairs in the antiparallel, parallel, and the $^{31}$P$^+$-P$_\mathrm{b0}^-$ state is denoted by [ap], [p] and [p+], respectively. }
\label{fig:Readout}
\end{figure}

In this section, we describe how the spin state of the $^{31}$P$_\mathrm{e}$-P$_\mathrm{b0}$ electron-spin pair in the absence of illumination is converted to a charge by means of an electron spin echo \cite{PRL100_177602} followed by a light pulse. This scheme is used for the measurements of the transition times $\tau_\mathrm{ap}$ and $\tau_\mathrm{p}$ of the spin pair into the $^{31}$P$^+$-P$_\mathrm{b0}^-$ state, shown in Fig.~1 of the main text.
The employed pulse sequence is shown in Fig.~\ref{fig:Readout} of this supplementary information; we use a 2-step phase cycle \cite{Schweiger_Book}, where the phase of the last $\pi$/2-pulse of the electron spin echo is changed by 180$^\circ$ with respect to the other pulses for every other sequence. Thus, the spin echo forms an effective 2$\pi$ pulse for sequence 1 and an effective $\pi$ pulse for sequence 2 \cite{PRL106_187601}. After a waiting time $T$, the echo is followed by an LED pulse during which the photocurrent through the sample is recorded by a fast digitizer card and boxcar integrated over typically 15~$\mu$s, resulting in a charge $\Delta Q$. Sequence 1 and sequence 2 are subtracted from each other and the result is averaged over 20-500 repetitions of the experiment.

We will discuss the dynamics of the $^{31}$P$_\mathrm{e}$-P$_\mathrm{b0}$ electron-spin pair in terms of the three states depicted in Fig.~1 (a) (i), (iii), and (iv) in the main text. We denote the fraction of spin pairs in the parallel spin state by [p], in the antiparallel spin state by [ap], and in the $^{31}$P$^+$-P$_\mathrm{b0}^-$ state by [p+], as sketched in Fig.~\ref{fig:Readout}. Assuming that at the beginning of the pulse sequence at the time (1) there are [ap]=$x$ antiparallel spin pairs and [p]=$y$ parallel spin pairs, a spin echo forming an effective $2\pi$ pulse results in [ap]=$x$ and [p]=$y$ at time (2). During the time interval $T$, chosen such that $\tau_\mathrm{p}\gg T\gg \tau_\mathrm{ap}$, all antiparallel spin pairs are transferred into the $^{31}$P$^+$-P$_\mathrm{b0}^-$ state, while the parallel spin pairs essentially remain unchanged, resulting in [ap]=0, [p]=$y$ and [p+]=$x$ at time (3). After switching on the light, a current transient occurs. Its spin-dependent part reflects the recombination of newly generated $^{31}$P$_\mathrm{e}$-P$_\mathrm{b0}$ spin pairs and the spin-dependent amplitude is therefore proportional to [p+]=$x$, the number of antiparallel spin pairs at time (1) before the detection pulse sequence. Repeating the same pulse sequence with a spin echo forming an effective $\pi$ pulse results in a current transient with its spin-dependent amplitude proportional to [p+]=$y$, the number of parallel spin pairs at time (1), as shown in Fig.~\ref{fig:Readout}. A large portion of the photocurrent transient, induced by the onset of the LED, is spin-independent and thus is independent of the phases of the applied microwave pulses; it is removed when sequences 1 and 2 are subtracted from each other. Thus a charge $\Delta Q$ is obtained as described above, which is proportional to the difference between the number of antiparallel and parallel spin pairs before the echo sequence, as stated in the main text. If the detection echo is preceded by an $\pi$ pulse, as depicted in Fig.~1~(b) of the main text, most spin pairs are in the antiparallel configuration at the time (1) and the lifetime of the antiparallel spin pairs $\tau_\mathrm{ap}$ can be determined.

In the case of the experiment to measure $\tau_\mathrm{p}$ of the parallel spin pairs to the $^{31}$P$^+$-P$_\mathrm{b0}^-$ state, there are in good approximation no antiparallel spin pairs at the beginning of the pulse sequence (1). Thus, the charge measured by the sequence shown in Fig.~\ref{fig:Readout} ($T\gg\ \tau_\mathrm{ap}$) is proportional to the amount of parallel spin pairs at the time (1), which is fixed in this experiment. As $T$ is increased to larger times ($T\approx \tau_\mathrm{p}$) the signal decreases because the parallel spin pairs will also undergo a transition into the $^{31}$P$^+$-P$_\mathrm{b0}^-$ state. This allows us to experimentally access the timeconstant $\tau_\mathrm{p}$, as shown in Fig.~1 (c) in the main text.

\section{Spin-to-charge conversion under continuous wave illumination}\label{Appendix_C}

In this section, we describe how the spin state of the $^{31}$P$_\mathrm{e}$-P$_\mathrm{b0}$ electron-spin is converted to a charge by means of an electron spin echo with phase cycling under continuous wave illumination \cite{PRL106_187601}. This mechanism is employed for the electrical readout of the nuclear spin polarization in Figs.~2 and 3 the main body of the paper. The pulse sequence, labeled ``repopulation and readout'' is shown in Fig.~2 (a) of the main text. The photocurrent is recorded and boxcar integrated directly after the microwave pulse sequence, as in previous pulsed electrically detected magnetic resonance experiments \cite{PRB68_245105}.

Under illumination a steady state of the electron spin system is established where in good approximation all spin pairs are in parallel configuration. The echo forming an effective 2$\pi$ pulse leaves the spin state unaffected, while the echo forming a $\pi$ pulse converts the spin pairs into antiparallel configuration; this leads to recombination of these spin pairs, resulting in a quenching of the photocurrent. By subtracting the charge obtained for the subsequent cycles, the spin-dependent signal can be discriminated from the spin-independent background. Since the measurement is performed on one hyperfine transition, the signal is proportional to the number of spin pairs with the nuclear spin of the $^{31}$P in one particular state, allowing for the nuclear spin readout presented in the main text.

In analogy to the phase cycle employed in the electron spin echo, the same concept can be applied for the measurement of the nuclear spin echoes. Here, the phase of the last radio frequency pulse of the nuclear spin echo was cycled as shown in Fig.~3 of the paper and the readout was performed by a single microwave $\pi$ pulse. This minimizes the total number of pulses, thus reducing the influence of pulse imperfections, and increases the signal by a factor of two, because each cycle yields a signal (of opposite polarity) which is proportional to the number of spins in the particular nuclear spin state.

%

\end{document}